\input phyzzx
\input tables
%
\REF\BEN{%
C. M. Bender and T. T. Wu,
\journal Phys. Rev. &184 (69) 1231;
\journal Phys. Rev. Lett. &27 (71) 461;
\journal Phys. Rev. &D7 (73) 1620;
\journal Phys. Rev. Lett. &37 (76) 117.}
\REF\COL{%
See, for example, S. Coleman,
in {\sl The Why of Subnuclear Physics}, A. Zichichi ed.,
(Plenum, New York, 1979).}
\REF\COLL{%
J. C. Collins and D. E. Soper,
\journal Ann. Phys. &112 (78) 209;\nextline
G. Auberson, G. Mennessier and G. Mahoux,
\journal Nuovo Cimento &A48 (78) 1.}
\REF\KLE{%
W. Janke and H. Kleinert,
\journal Phys. Rev. &A42 (90) 2792;\nextline
H. Kleinert,
\journal Phys. Lett. &B300 (93) 261;
\journal Phys. Lett. &A190 (94) 131;
quant-ph/9507005;
\journal Phys. Lett. &B360 (95) 65;\nextline
R. Karrlein and H. Kleinert,
\journal Phys. Lett. &A187 (94) 133;\nextline
H. Kleinert and W. Janke,
\journal Phys. Lett. &A206 (95) 283;\nextline
H. Kleinert and S. Thoms,
\journal Phys. Rev. &D52 (95) 5926;\nextline
H. Kleinert and I. Mustapic,
\journal Int. J. Mod. Phys. &A11 (96) 4383;\nextline
H. Kleinert, S. Thoms and W. Janke,
quant-ph/9605033;\hfil\break
H. Kleinert, S. Thoms, and V. Schulte-Frohlinde,
quant-ph/9611050.}
\REF\SEZ{%
R. Seznec and J. Zinn-Justin,
\journal J. Math. Phys. &20 (79) 1398;\nextline
J. C. Le Guillou and J. Zinn-Justin,
\journal Ann. Phys. &147 (83) 57;\nextline
J. Zinn-Justin,
{\sl Quantum Field Theory and Critical Phenomena},
(Oxford, New York 1996);\nextline
see also, K. Hiraizumi, Y. Ohshima and H. Suzuki,
\journal Phys. Lett. &A216 (96) 117.}
\REF\GKS{%
R. Guida, K. Konishi and H. Suzuki,
\journal Ann. Phys. &241 (95) 152;
\journal Ann. Phys. &249 (96) 109.}
\REF\REVIEW{%
Many important works in this field are reprinted in
{\sl Large-Order Behavior of Perturbation Theory},
J. C. Le Guillou and J. Zinn-Justin eds.,
(North-Holland, Amsterdam, 1990).}
\REF\AOY{%
H. Aoyama, T. Harano, H. Kikuchi, M. Sato and S. Wada,
hep-th/9606159, and references therein.}
\REF\RIC{%
For a recent article on the application of Borel resummation, see,
R. Guida and J. Zinn-Justin, hep-th/9610223.}
\REF\LOE{%
J. J. Loeffel, Centre d'Etudes Nucl\'eaires de Sacley Report
DPh-T/76-20 (1976), reprinted in~[\REVIEW].}
\REF\PAR{%
G. Parisi,
\journal Phys. Lett. &B69 (77) 329.}
\REF\BRE{%
E. Br\'ezin, J. C. Le. Guillou and J. Zinn-Justin,
\journal Phys. Rev. &D15 (77) 1558;\nextline
E. Br\'ezin, G. Parisi and J. Zinn-Justin,
\journal Phys. Rev. &D16 (77) 408.}
\REF\ZIN{%
J. Zinn-Justin,
\journal Nucl. Phys. &B192 (81) 125;
\journal Nucl. Phys. &B218 (83) 333;
\journal J. Math. Phys. &25 (84) 549.}
\REF\LAU{%
B. Lautrup,
\journal Phys. Lett. &B69 (77) 109;\nextline
G. 't Hooft,
in {\sl The Why of Subnuclear Physics}, A. Zichichi ed.,
(Plenum, New York, 1979);\nextline
P. Olesen,
\journal Phys. Lett. &B73 (78) 327;\nextline
G. Parisi,
\journal ibid. &B76 (78) 65.}
\REF\BREZ{%
E. Br\'ezin and G. Parisi,
\journal J. Stat. Phys. &19 (78) 269.}
%
%
\overfullrule=0pt
\pubnum={IU-MSTP/17; hep-th/9612165}
\date={February 1997}
\titlepage
\title{Observing Quantum Tunneling in Perturbation Series}
\author{%
Hiroshi Suzuki\foot{%
e-mail: hsuzuki@mito.ipc.ibaraki.ac.jp}
and
Hirofumi Yasuta\foot{%
e-mail: yasuta@mito.ipc.ibaraki.ac.jp}
}
\address{%
Department of Physics, Ibaraki University, Mito 310, Japan}
\abstract{%
We apply Borel resummation method to the conventional perturbation
series of ground state energy in a metastable potential,
$V(x)=x^2/2-gx^4/4$. We observe numerically that the discontinuity of
Borel transform reproduces the imaginary part of energy eigenvalue,
i.e., total decay width due to the quantum tunneling. The agreement
with the exact numerical value is remarkable in the whole tunneling
regime $0<g\lsim0.7$.}
\endpage
The quantum tunneling is a purely non-perturbative phenomenon: This
phrase has been widely accepted. Let us consider a simple quantum
mechanical example
$$
   H={p^2\over2}+{1\over2}x^2-{g\over4}x^4.
\eqn\one
$$
Since the ground state is metastable in this potential, the eigenvalue
is defined by the analytic continuation from~$g<0$. Equivalently, the
Schr\"odinger equation is defined with a rotated boundary condition,
$\psi(x)\to0$ for $x=e^{\pi i/6}t$ with $t\to\pm\infty$~[\BEN],
if the ``escape out'' solution is taken. With this
boundary condition, the semi-classical (WKB) calculation~[\BEN] gives
the imaginary part of energy eigenvalue for~$g\ll1$,
$$
   {\rm Im}\, E(g)\sim-\sqrt{8\over\pi g}
   \exp\left(-{4\over3g}\right)\left[1-{95\over96}g+O(g^2)\right].
\eqn\two
$$
This is related to the total decay width~${\mit\Gamma}$ due to the
quantum tunneling,\foot{%
We may characterize the tunneling regime of the coupling constant
space as the real part of energy is lower than the potential barrier
height. The numerical solution shows~$0<g\lsim0.677$ is the
tunneling regime.}
${\mit\Gamma}=-2\,{\rm Im}\,E(g)$. The semi-classical result~\two\
vanishes to all order of the expansion on~$g$, i.e., the
tunneling effect is invisible in a simple expansion with respect to
the coupling constant.

On the other hand, the conventional Rayleigh--Schr\"odinger
perturbation series of ground state energy is given by
$$
   E(g)\sim\sum_{n=0}^\infty c_ng^n,
\eqn\three
$$
where the first several coefficients read\foot{%
We have prepared the exact perturbative coefficients $c_n$ to~$n=153$,
using the technique in~[\BEN].}
$$
   c_0=1/2,\quad c_1=-3/16,\quad c_2=-21/128,\quad c_3=-333/1024,
   \quad\cdots.
\eqn\four
$$
For $g$~real, the simple truncated sum of the expansion~\three\ is of
course real and does not produce the desired imaginary part.
Therefore usually one needs some non-perturbative technique, such as
the semi-classical approximation of Schr\"odinger equation or, of the
Euclidean path integral, to estimate the tunneling amplitude. In the
language of the Euclidean path integral, the exponent in~\two\ is
given by the action of the bounce solution~[\COL] and the $O(g)$~term
is the two loop radiative correction around the bounce~[\COLL].

In this article, we re-examine the above common wisdom. We try to
reproduce the imaginary part by {\it solely\/} relying on the
conventional perturbation series~\three. This kind of approach has
been pursued by Kleinert and co-workers recently~[\KLE]. In their
work, a variational perturbation method, which is equivalent to
the order dependent mapping (ODM) method~[\SEZ], was applied to resum
the perturbation series. They showed numerically that an accurate
value of the imaginary part was reproduced in the strong coupling
regime~$g\gsim0.1$. Later, the rigorous convergence proof of ODM
was given~[\GKS] for $g>g_0\simeq0.1$, where $g_0^{-2/3}$ is the
convergence radius of the strong coupling expansion.

ODM works quite well in the strong coupling regime. However it cannot
take the place of the semi-classical approximation because it fails
in the weak coupling regime~$g\lsim0.1$. This is not satisfactory
from the above motivation although the failure can be understood~[\GKS]
as the effect of Bender--Wu singularity~[\BEN] on the higher Riemann
sheet of the complex $g$~plane. As the another undesirable feature
of ODM, the answer is given as a root of a higher algebraic equation,
hence does not allow a simple analytical characterization such
as~\two.\foot{%
H.S. thanks T. Eguchi for his notice on this point: This was the
original motivation of the present work.}

We nevertheless do not believe the above disadvantage of ODM itself is
the fundamental difficulty of an approach based on the perturbation
series. In fact, from the extensive studies in the
seventies~[\REVIEW], it has been known that the large order behavior
of perturbation series is determined by the tunneling effect (at
least in super-renormalizable cases). Therefore it is rather natural
to expect the leading semi-classical behavior~\two\ for~$g\ll1$ can
be extracted from the information of perturbation series. The idea
behind our approach is simple: If the {\it large\/} order behavior of
perturbation series is determined by the lowest semi-classical
contribution~\two, the {\it lower\/} order perturbation coefficients,
which we {\it can\/} compute reliably, should contain the information
on the higher order corrections. We shall see below this expectation
is in fact correct.

This possibility is important even practically because the systematic
higher order correction to the lowest semi-classical calculation is
not a simple matter, especially in many variable systems: One should
include the correction due to multi-bounce type configuration, with an
integration over the (quasi-)collective coordinates, and the
perturbative expansion around the multi-bounce, and so on. Such a
problem, even in quantum mechanics, is still under active current
researches (see, for example~[\AOY]). On the other hand, the
perturbative calculation is the best-established technique in
quantum theory.

Here we take a classical and conservative approach to this problem.
Namely we apply to the perturbation series the Borel resummation
method~[\REVIEW,\RIC], which proceeds as follows: We first define
Borel transform from the perturbative coefficients~$c_n$ in~\three,
$$
   B(z)\equiv\sum_{n=0}^\infty{c_n\over n!}z^n.
\eqn\five
$$
Then the energy eigenvalue is defined by Borel integral
$$
   E(g)\equiv{1\over g}\int_0^\infty dz\,e^{-z/g}B(z).
\eqn\six
$$
However the imaginary part of energy eigenvalue implies the Borel
transform develops a singularity on the positive real axis. The
imaginary part~\two\ for~$g\ll1$ is reproduced by a fractional branch
point
$$
   B(z)\sim f(z_0)(z_0-z)^\alpha+\cdots,
\eqn\seven
$$
where $f(z_0)=-2\sqrt{2}/\pi$, $z_0=4/3$ and $\alpha=-1/2$; this is
the singularity nearest to the origin and there may exist other
singularities. The integration along the positive real axis~\six\ is
therefore ill-defined as it stands (the so-called ``non-Borel
summable'' case) and we define the integral by deforming the
integration contour to the {\it upper\/} side of branch cut.
From the final result, we shall see this choice of contour
corresponds to our boundary condition in the original Schr\"odinger
problem.

Incidentally, to find the position of the nearest singularity of Borel
transform~$z_0$, we have used the information of semi-classical
calculation~\two. In principle, those values of $z_0$ and~$\alpha$
may be found solely from the perturbation series~$c_n$. Assuming only
the first term of~\seven, we have a relation
$$
\eqalign{
   &{(n+1)c_n\over c_{n+1}}={n+1\over n-\alpha}z_0,
\cr
   &-n^2\left[{(n+1)c_n\over c_{n+1}}-{nc_{n-1}\over c_n}\right]
   {1\over z_0}-1
   ={\alpha+1\over1-(2\alpha+1)/n+\alpha(\alpha+1)/n^2}-1,
\cr
}
\eqn\eight
$$
whose $n\to\infty$ limit is nothing but the Appel's comparison
theorem. We found numerically that $z_0=1.33341$
and~$\alpha=-0.518130$ for $n=152$, which are more or less consistent
with the exact values. After observing this, we will use the exact
value~$z_0=4/3$ in what follows.

What we have to do is a construction of Borel transform~\five\
from the perturbation coefficients~$c_n$ and Borel integration~\six\
with the deformed contour. Unfortunately this simple recipe does not
work practically because the convergence radius of series~\five\ is
finite~($=z_0$) due to the branch point singularity~\seven. To perform
the integration along the positive real axis, we have to continue
analytically the series~\five\ outside the convergence circle, which is
impossible without having all the coefficients of the series. As is
well known, however, the analytic continuation can be avoided by
the conformal mapping technique~[\LOE].

We introduce a new variable $\lambda$ by
$$
   z=4z_0{\lambda\over(1+\lambda)^2}.
\eqn\nine
$$
This transformation maps the whole cut $z$~plane into an interior of a
unit circle on $\lambda$~plane. In particular, an interval
$z\in[0,z_0)$ is mapped to $\lambda\in[0,1)$ and,
$z\in[z_0+i\varepsilon,\infty+i\varepsilon]$ is to the upper arc of a
circle with radius~$1-\varepsilon$ on $\lambda$~plane. The infinity
$z=\infty$ is mapped to $\lambda=-1$. In terms of~$\lambda$, the
series~\five\ is expressed as
$$
   B(z)=\sum_{k=0}^\infty d_k\lambda^k,\quad
   d_k\equiv\sum_{n=0}^k(-1)^{k-n}
   {\Gamma(k+n)\over(k-n)!\,\Gamma(2n)}
   (4z_0)^n{c_n\over n!}.
\eqn\ten
$$
The point is that, assuming the absence of singularity of~$B(z)$
on the cut $z$~plane,\foot{%
The branch point of $B(z)$ at $z=z_0$ is transformed to a single pole
at~$\lambda=1$. We investigated other possible singularities of
$B(z)$ on $\lambda$~plane by applying to the series~\ten\ the
Appel's comparison theorem~\eight\ ($c_n\to d_n$). It does not
converge to a definite value and indicates the other singularities
{\it on\/} a positive real axis of $z$~plane $z>z_0$. We suspect this
is an effect of the multi-bounce configuration.}
the convergence circle of the series~\ten\ now is the unit circle,
within which the whole cut $z$~plane is mapped. Therefore we may use
a truncated sum of the expansion~\ten\ in the integration~\six. This
trick has been extensively used in practical applications of Borel
resummation~[\REVIEW,\RIC]. However, in our present case, one
subtlety arises: We have to integrate~\ten\ {\it along\/} its
convergence circle, on which the (uniform) convergence of the series
is not guaranteed in general. For us, this seems to be one reason why
the Borel resummation has not been seriously applied to the tunneling
phenomenon.\foot{%
The application of Borel resummation to~\three\ with $g<0$ (anharmonic
oscillator) is found in~[\PAR].}
Anyway we can now go on, hoping the convergence of the method.

We then parameterize the unit circle as~$\lambda=e^{i\theta}$.
From~\ten, \nine\ and~\six, we find that the $N$th order
approximation of the imaginary part is given by
$$
   \left[{\rm Im}\,E(g)\right]_N
   ={z_0\over g}\int_0^\pi d\theta\,
   \exp\left(-{z_0\over g}{1\over\cos^2\theta/2}\right)
   {\sin\theta/2\over\cos^3\theta/2}
   \sum_{k=0}^N d_k \sin k\theta,
\eqn\eleven
$$
where $d_k$ is defined in~\ten. Note that this is solely expressed by
the perturbative coefficients~$c_n$ to $N$th order (except the value
of~$z_0$ for which we may use the semi-classical method, or
$n\to\infty$ limit of~\eight). Therefore this is the analytic
expression of tunneling amplitude which we were looking for. In fact,
a contribution from an infinitesimal interval near $\theta\sim0$
in~\eleven\ is proportional to $\sim\exp(-z_0/g)$, the semi-classical
behavior. Remaining integration may be regarded as its higher order
corrections.\foot{%
It might look strange that we have a finite imaginary part from the
integration on $\lambda$ plane, because an integration along the real
axis $-1<\lambda\le1$ (which does not give the imaginary part) may be
added to close the integration contour, and we have assumed the
absence of singularity inside the unit circle. In fact there is no
contradiction because the change of variable~\nine\ produces in the
integrand a double pole at $\lambda=-1$ and a infinitesimal $-\pi/2$
rotation around $\lambda=-1$ can produce minus of the imaginary part.}

In Fig.~1, we plotted the logarithm of the relative error of~\eleven\
to the exact imaginary part for several values of~$g$. We
observe a very rapid convergence and practically we may even use
$N\sim5$ which only gives a few percent error for $g\sim0.3$!
In Fig.~2, eq.~\eleven\ with $N=15$ is plotted as a function of~$g$.
For a comparison, the exact numerical value and the semi-classical
calculation including the two loop correction~\two\ are also plotted.
In Table~1, several values in the weak coupling regime are listed.
The agreement with the exact value in a whole tunneling
regime~$0<g\lsim0.7$ is remarkable and, as we expected, it is
much better than the semi-classical result: It provides
a systematical quantitative improvement of the bounce calculation.

We have seen that for the metastable potential~\one, the
(generalization of) Borel resummation method combined with the
conformal mapping technique reproduces the {\it very\/} accurate
tunneling amplitude. We expect a similar result holds for generic
{\it metastable\/} potentials, for which the imaginary part is
``physical.'' On the other hand, the tunneling phenomenon also exists
in a {\it bounded\/} potential but with multi-minima (the quantum
coherence). In this case also the perturbation series is non-Borel
summable due to the tunneling~[\BRE] and the Borel resummation
produces imaginary part. However since the potential is bounded, this
cannot be regarded as a physical imaginary part\foot{%
This unphysical imaginary part should be canceled by an
imaginary part arising from the quasi-collective coordinate integration
of multi-instantons~[\ZIN].}
and our approach, based on a conventional perturbation series
around the ``trivial vacuum,'' cannot directly be applied to such a
bounded potential. The imaginary part will merely give the order of
magnitude of barrier penetration, or typical time scale of an
oscillation among different minima.

Finally we comment on the generalization to higher dimensional models.
The quantum mechanics~\one\ we have analyzed is of course so simple
that everything can be computed numerically. However it is
equivalent to a one-dimensional ($D=1$) scalar field theory:
$$
   S=\int d^Dx\left({1\over2}\partial_\mu\phi\partial^\mu\phi-
                      {1\over2}\phi^2+{g\over4}\phi^4\right).
\eqn\twelve
$$
We can expect a similar set of formulas as \ten\ and~\eleven\
(for the energy density) works for~\twelve, if~$D<4$ or with an
ultraviolet cutoff. For these cases, the model is super-renormalizable
or ultraviolet finite and thus the renormalon~[\LAU], another known
source of the Borel singularity, does not emerge. An analysis
of the Euclidean bounce solution shows that the nearest singularity
is, $z_0=5.850448$ for $D=2$ and $z_0=18.897251$ for $D=3$, with
$\alpha=-D/2$~[\BREZ]. What we have to compute is (appropriately
renormalized) vacuum bubble diagrams as much higher orders as
possible. Then an accurate bubble nucleation rate in a certain
$D-1$~dimensional system will be reproduced. A study along this line
is in progress.

The work of H.S. is supported in part by the Ministry of Education
Grant-in-Aid for Scientific Research, Nos.~08240207, 08640347,
08640348 and~07304029.

\refout
\vfill\eject
\centerline{\fourteenrm Figure Captions}
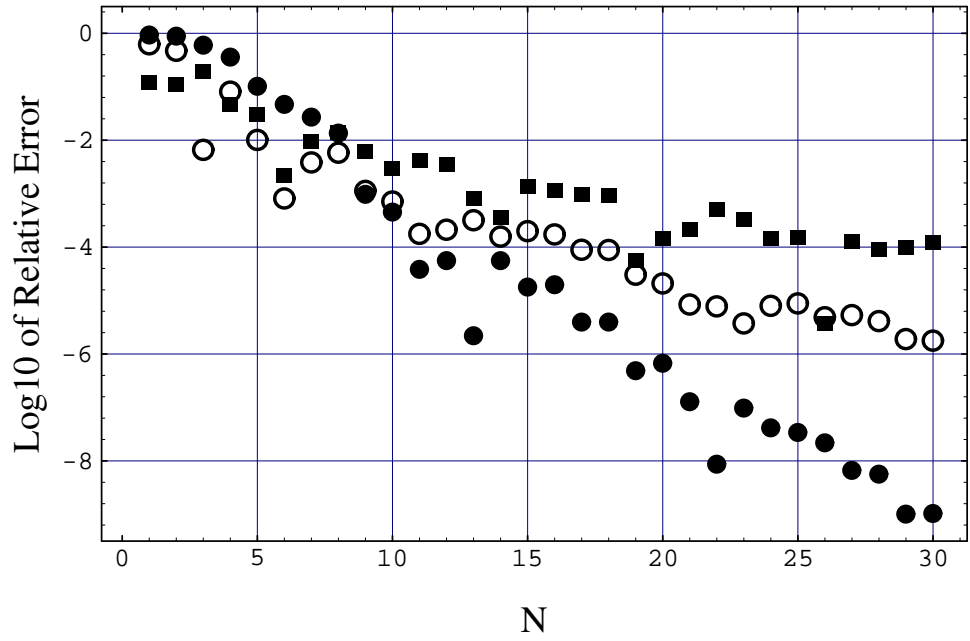
\item{\rm Figure\ 1.}
The logarithm of the relative error of~\eleven\ to the exact value
for, $g=0.08$ (the full circle), $g=0.3$ (the circle) and $g=0.6$
(the square), as a function of~$N$.
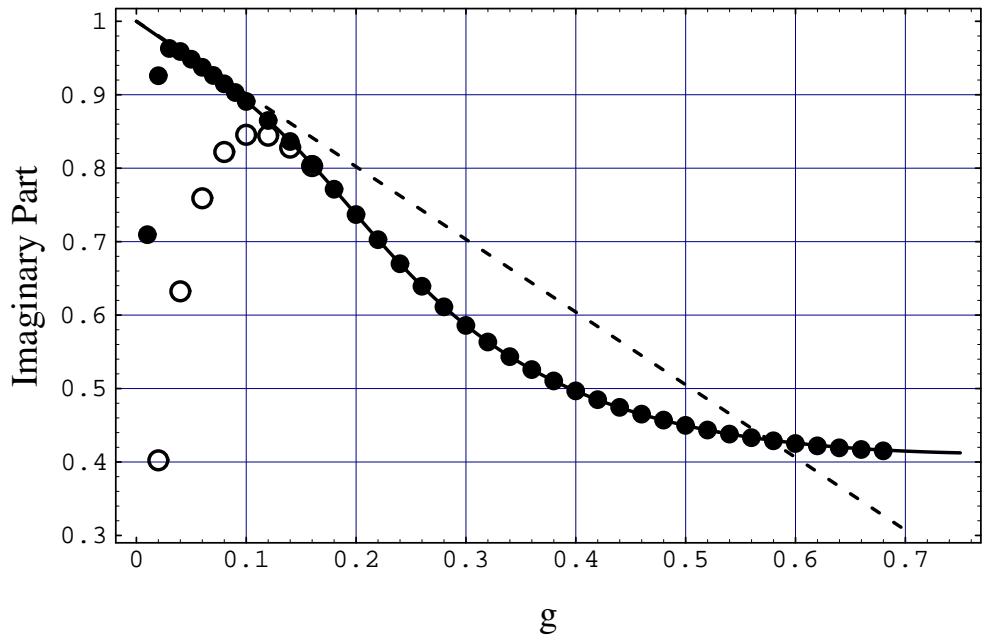
\item{\rm Figure\ 2.}
The imaginary part (normalized by the lowest semi-classical
result, the first term of \two) computed by~\eleven\ with~$N=15$
(the full circle). For a comparison, the exact numerical value
(the solid line) and the semi-classical result~\two\ (the broken line)
are plotted. To indicate the convergence behavior, the result
with~$N=5$ is also plotted for $g\leq0.16$ by circles.
\bigskip
\centerline{\fourteenrm Table Caption}
\item{\rm Table\ 1.}
Comparison of \eleven\ ($N=15$) and the exact numerical value of
imaginary part in the weak coupling regime. All the numbers are
normalized by the lowest semi-classical result, the first term
of~\two.
\vfill\eject
\null
\vfill
\begintable
$g$ \| exact | semi-classical (eq.~\two) | eq.~\eleven\ with $N=15$
\crthick
\quad$0.08$\quad\hfill \| \quad$0.914996$\quad | $0.920833$ |
$0.915012$
\cr
\quad$0.1$\quad\hfill \| \quad$0.891039$\quad | $0.901042$ |
$0.891023$
\cr
\quad$0.12$\quad\hfill \| \quad$0.865072$\quad | $0.881250$ |
$0.865082$
\endtable
\vfill
\centerline{\fourteenrm Table 1}

\bye

\null
\vfill
\centerline{
\vbox{
\halign{%
\quad#\hfil\quad&\quad#\hfil\quad&\quad#\hfil\quad&\quad#\hfil\quad
\cr
$g$ & exact & semi-classical (eq.~\two) & eq.~\eleven\ with $N=15$
\cr
$0.08$ & $0.914996$ & $0.920833$ & $0.915012$
\cr
$0.1$ &  $0.891039$ & $0.901042$ & $0.891023$
\cr
$0.12$ & $0.865072$ & $0.881250$ & $0.865082$
\cr}
}
}
\vfill
\centerline{\fourteenrm Table 1}
\bye